\documentclass[fleqn,twoside]{article}
\usepackage{espcrc2}

\usepackage{graphicx}
\usepackage[figuresright]{rotating}

\newcommand{\AmS}{{\protect\the\textfont2
  A\kern-.1667em\lower.5ex\hbox{M}\kern-.125emS}}
\newcommand{\eq}[1]{{\frenchspacing Eq.~(\ref{#1})}}
\newcommand{\fig}[1]{{\frenchspacing Fig.~(\ref{#1})}}

\newcommand{\beq}{\begin{equation}}
\newcommand{\eeq}{\end{equation}}
\newcommand{\ep}{$\eta'$}
\newcommand{\bc}{\begin{center}}
\newcommand{\ec}{\end{center}}
\title{Flavour Singlet Mesons in full QCD on the Lattice}

\author{K. Schilling\address[WUPP]{Fachbereich Physik,
    Bergische Universit\"at, 
    D-42097 Wuppertal, Germany }, H. Neff\addressmark[WUPP]\thanks{since Oct
    1, 2001: IASA, Athens, Greece},
    N. Eicker\addressmark[WUPP]\thanks{since Dec 1, 2001: ParTec AG, Karlsruhe}, Th.
  Lippert\addressmark[WUPP],
        J.W. Negele\address{Center for Theoretical Physics, MIT, 77 Massachusetts Ave,
    Cambridge, MA 02139, USA    }
 }
       
\begin{document}

\begin{abstract}
  We apply spectral methods to compute the OZI-rule suppressed loop-loop
  correlators in the pseudocalar meson flavour singlet channel. Using SESAM
  configurations (obtained with two degenerate sea quark flavours on
  $16^3\times 32$ lattices at $\beta = 5.6$, with standard Wilson action), we
  find for the first time clear evidence for mass plateau formation in the \ep
  channel of this theory.  As a consequence, we observe a clear signal of a
  mass gap persistent under chiral extrapolation.  This sets the stage for a
  more realistic two-channel approach, where partially quenched strange quarks
  would be included, in addition to the $u,d$ sea quarks.

  \vspace{1pc}
\end{abstract}

\maketitle
\section{INTRODUCTION}
The large empirical mass gap between the $\eta'$ and $\pi$ masses, $m_0^2 =
m_{\eta'}^2 - m_{\pi}^2 $, is commonly believed as being due to the $U_A(1)$
anomaly of the flavour singlet axial vector current. In this picture, the
large $\eta'$-meson mass is driven by quark-antiquark interactions mediated by
topological charge fluctuations in the QCD vacuum. In the $N_c = \infty$-limit
of QCD this scenario is put forward through the famous
Witten-Veneziano~\cite{witten} formula (WVF) that models $m_0^2$ in terms of
the topological susceptibility, from quenched QCD. The pitfalls of a lattice
evaluation of the Witten-Veneziano formula have been elucidated in
Ref.~\cite{rossi}.

In principle an {\it ab initio}, direct  verification of the $U_A(1)$ origin of flavour
singlet pseudoscalar masses {\it beyond the large $N_c$-approximation} can be
accomplished by simulating unquenched QCD on the lattice, but conclusive results
on the mass gap have been impeded to date~\cite{ecfa}, mainly due to severe
signal-to-noise problems in the analysis of the flavour singlet
correlator~\cite{ukqcd,cppacs,sesam}.

\begin{figure}[th]
\vspace{9pt}
\centerline{
\includegraphics[width=50mm]{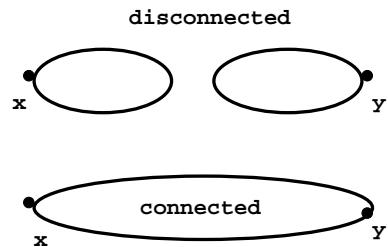}}
\vspace{-10mm}
\caption{The  two contributions to the $\eta'$ propagator {\it in the full} QCD vacuum:
 the connected part with running through  valence quark
  lines and the `disconnected'  piece where the valence
  quark lines annihilate instead of connecting source $x$ and sink $y$ at time
slices $t=0$ and $t =\Delta t$. }
\label{fig:picto}
\end{figure}

\begin{figure}[th]
\vspace{9pt}
\centerline{\includegraphics[bb = 50 50 409 303, width = .5\textwidth]{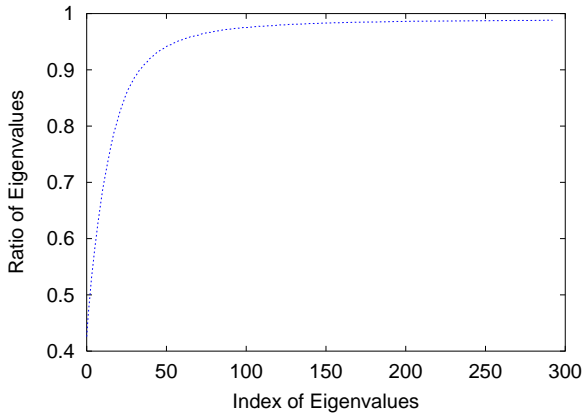}} 
\vskip .5cm
\vspace{-10mm}
\caption{The net variation  of  low eigenmodes of $Q$
  over the sea quark mass range reached by the SESAM simulation, as expressed
  in the spread of  eigenmode ratios, $R = |\lambda_i(0.1575) /
  \lambda_i(0.1560)|$, with the eigenvalues  ordered according
to increasing moduli.}
\label{fig:spectrum}
\end{figure}
\begin{figure}[th]
\vspace{9pt}
\centerline{\includegraphics[bb = 50 50 409 303, width = .5\textwidth]{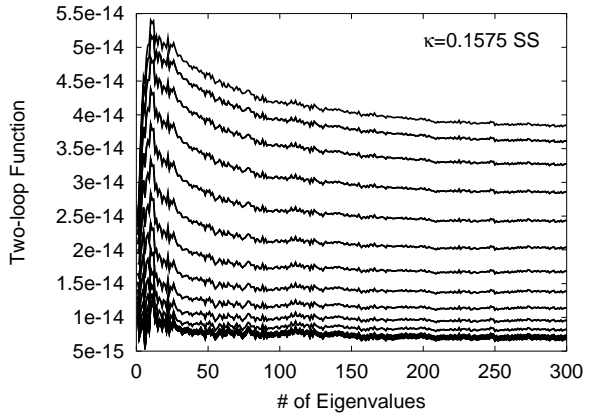}} 
\vskip .5cm
\vspace{-10mm}
\caption{The spectral cutoff dependence 
of the disconnected contribution, $D^l(\Delta t)$, for various
time separations $\Delta t = 1 , \cdots 16$ (top to bottom), at the lightest
SESAM sea quark mass. }
\label{fig:looploop}
\end{figure}
\begin{figure}[thb]
\vspace{9pt}
\centerline{\includegraphics[bb = 50 50 409 303, width = .5\textwidth]{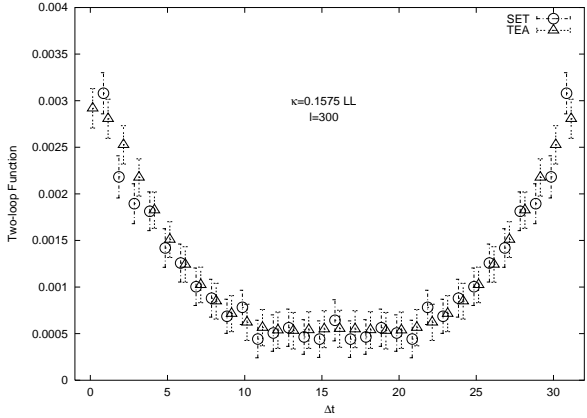}} 
\vskip .5cm
\vspace{-10mm}
\caption{
The disconnected contribution to the \ep -correlator from
its spectral representation with cutoff $l=300$, at the
smallest sea quark mass of the SESAM simulation.}
\label{fig:disconnected}
\end{figure}
\begin{figure}[thb]
\vspace{9pt}
\centerline{\includegraphics[bb = 50 50 409 303, width = .5\textwidth]{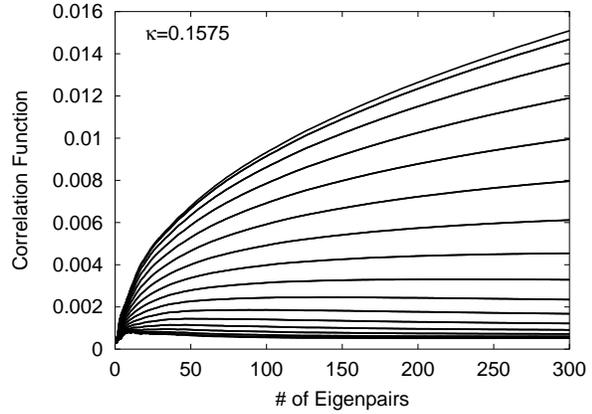}} 
\vskip .5cm
\vspace{-10mm}
\caption{
The analogue to Fig. (3) , but for the connected
contribution, $ C(\Delta t)$, to Eq. (3).}
\label{fig:oneloop}
\end{figure}
\begin{figure}[bht]
\vspace{9pt}
\centerline{\includegraphics[bb = 50 50 409 303, width = .5\textwidth]{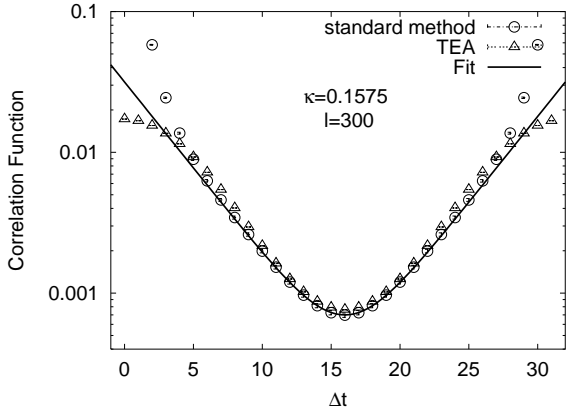}} 
\vskip .5cm
\vspace{-10mm}
\caption{
The one loop correlator computed by standard linear solver techniques (labeled
`standard method') and TEA with spectral cutoff $l=300$, at the lightest SESAM sea quark mass. Also
shown is the ground state projection (marked `Fit').}
\label{fig:oneloop1}
\end{figure}

\begin{figure}[thb]
\vspace{9pt}
\centerline{\includegraphics[bb = 50 50 409 303, width = .5\textwidth]{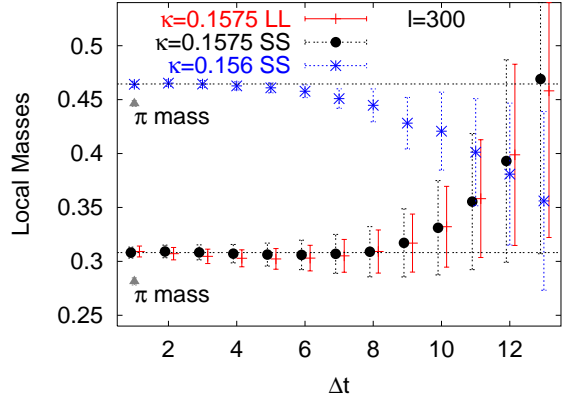}} 
\vskip .5cm
\vspace{-10mm}
\caption{
  Effective \ep -mass in lattice units at the lightest ($\kappa = 0.1575$) and
  heaviest ($\kappa = 0.1560$) sea quark masses of the SESAM simulation,
both with (SS) and without (LL) source and sink smearing~\cite{smeared}.}
\label{fig:effmass}
\end{figure}
\begin{figure}[thb]
\vspace{9pt}
\centerline{\includegraphics[bb = 50 50 409 303, width = .5\textwidth]{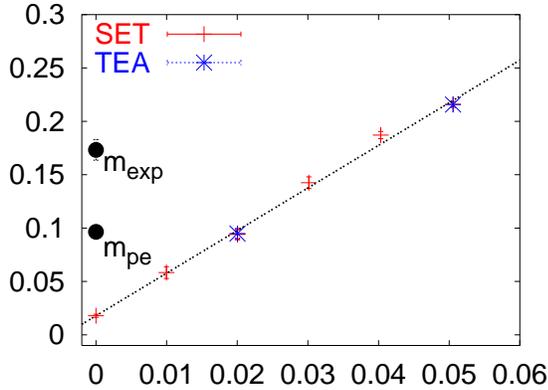}} 
\vskip .5cm
\vspace{-10mm}
\caption{
  Demonstration of the quality of data: chiral extrapolation of $m_{\eta'}^2$,
  plotted {\it vs.}  sea quark mass $m_{\mbox{sea}}$, all in lattice units.
$m_{pe}$ is a naive estimate from WVF (replacing $m_0^2 \rightarrow 2/3
  m_0^2$ to accomodate to $N_f=2$).}
\label{fig:extrapol}
\end{figure}

As shown in \fig{fig:picto}, the correlator in question, 
 \beq
C_{\eta'}(\Delta t) = <0|{\cal O}^{\dagger}(\Delta t){\cal O} (t =0)|0> \eeq
contains  
a loop-loop (fermionically
`disconnected') contribution, $D$.
This OZI-suppressed piece  arises from
the Wick contractions within  each of the two light quark bilinears \beq
{\cal O} = \bar{q}\gamma_5 q 
\eeq 
and was evaluated in the literature by use of stochastic estimator
techniques (SET), to cover its entire volume dependence.
As a net result
the $\eta'$ propagator $C_{\eta'}(\Delta t)$ is computed as the {\it numerical
  difference} between its connected and disconnected  pieces:
\beq
C_{\eta'}(\Delta t) = C(\Delta t) - 2 D(\Delta t) \stackrel{\Delta t \to \infty}{\longrightarrow}  \mbox{exp}
 (-m_{\eta'}\Delta t)\; ,
\label{eq:etacorr}
\eeq and the \ep -mass is to be extracted from its asymptotic behaviour.  
Since the
large mass $m_{\eta'}$ corresponds to a strong numerical cancellation
between the positive terms $C$ and $D$, the calamity of previous LQCD attempts
was to establish a window in $\Delta t$ with unambiguous single exponential
decay, before signals disappear under the noise level.

Barring brute force methods it
 appears that important progress in the field could  be achieved by 
\begin{enumerate}
\item
 avoiding SET  and/or 
\item
accomplish    precocious ground state dominance
in the flavour singlet channel.
\end{enumerate}

With this motivation, we report here on some recent methodological studies of
ours~\cite{tea,neff_diss} where we readdress the unsettled problem of the
direct $\eta'$ signal in LQCD and explore the possibility of spectral representations
of quark propagators in superseding the SET treatment of $D(\Delta
t)$~\cite{ukqcd,cppacs,sesam}.  Such approach is of course in lign with the
expectation that chiral physics should be controlled by the low modes of the
Dirac-Wilson operator.  The point to find out is whether in the sea quark mass
range of  state-of-the-art {\it full} QCD simulations -- like the
 SESAM simulation~\cite{spectrum} -- a reasonably small number of low
eigenmodes actually suffices to saturate the fermion loop expansions with
appropriate degree of accuracy.

\section{The spectral approach}
Since the Wilson-Dirac matrix, $M$, is  non-normal at non-zero lattice spacings,
 it is preferable to  expand in
the eigenspectrum of its Hermitian equivalent
\beq Q^{\dagger} = Q :=\gamma_5 M\; .
\label{eq:hermi}
\eeq 
$Q$ has been shown to provide the optimal eigenvalue basis in the sense
of the singular value decomposition~\cite{hip}.  

For our practical benchmarking we use 200 (195) SESAM lattices of size $16^3
\times 32$ with $\kappa = 0.1575$ ($0.1560$)~\cite{spectrum,smeared}.
A detailed account  of our analysis can be found in  Ref.~\cite{neff_diss}.

The eigenmodes of $Q$ can be determined by means of Lanczos methods.  In
\fig{fig:spectrum} the 300 lowest eigenvalues are plotted to show their
response under the SESAM variation in sea quark mass.  The characteristic
range of this response in eigenvalue space suggests that a truncated eigenmode
approach (TEA),
based on some  {\it few hundred low modes},
might indeed  encompass the infrared physics of
interest.

In terms of the
eigenstates of $Q$, 
\beq Q | \psi_i \rangle = \lambda_i | \psi_i \rangle \; ,
\label{eq:eigen}
\eeq
the quark propagator can  readily be computed, from any source $z$
to every sink $z'$:
\begin{eqnarray}
&M^{-1}(z,z') &= \nonumber\\
&&\sum_{i = 1}^{{ l}}\frac{\gamma_5}{\lambda_i} 
\frac{ | \psi_i (z) \rangle \langle \psi_i (z') | }
{ \langle \psi_i |  \psi_i \rangle}\; .\label{eq:inverse}
\end{eqnarray}
From this expression it is straightforward to compute the loop-loop correlator
of \fig{fig:picto} in spectral representation, $T$; it should be dominated by
a small number of modes at large enough values of $\Delta t$.  This is indeed
reflected in \fig{fig:looploop} where we display the $l$-dependence of the
disconnected contribution to the \ep -propagator, $D^l$, the family of curves
representing the set of different time slices, $\Delta t = 1, \cdots, 16$
(from top to bottom).  While we find a few tens of modes to bear out the long
range (in $\Delta t$) features of $D$, a few hundred modes appear to saturate
its short range structure under the SESAM conditions. This finding is also
visualized in \fig{fig:disconnected} where we plotted $D^{300}(\Delta t)$
together with the previous SET result~\cite{sesam} and find them to agree
nicely within their errors.

For the connected piece of the \ep -correlator, $C$, the viability of TEA is
less evident as can be seen from \fig{fig:oneloop}. Even with a cutoff
$l=300$, saturation at large $\Delta t$ is not fulfilled in detail, not to
speak of the region of small time separations, where excited state
contributions are not covered at all by the 300 lowest modes. This is of
course not an obstacle for our present purposes, since this part of the
correlator can easily be treated by standard methods.  In fact, at this point,
we go one step beyond and {\it exploit the high accuracy} of $C$ as achieved
from iterative solvers: we replace $C$ by its ground state contribution, $C^g_{\pi}$
(as obtained from a single exponential fit at large time separations, see
\fig{fig:oneloop1}), $C \rightarrow C^g_{\pi}$. This then leads us to the final form
of the projected \ep -correlator
\begin{equation}
\label{eq:neffs}
\tilde{C}_{\eta'}(\Delta t)={ C^g_{\pi}(\Delta t)}- 2 T(\Delta t) \; . 
\end{equation}
in terms of  the two loop spectral approximation, $T$:
\beq
\label{eq:tlf}
T(\Delta t) = \mbox{tr}|_{t=0}\; Q^{-1} \quad \mbox{tr}|_{\Delta t}\;
Q^{-1}\; , \eeq 
with $N_f =2$ being the number of active sea quark flavours and
$Q^{-1}$ taken from \eq{eq:inverse}.

This enables us to carry out a standard effective local \ep -mass analysis,
which is much more sensitive than looking at correlators or ratios
thereof~\cite{cppacs}.  As a result we find an effective mass plot
(see \fig{fig:effmass}) with striking mass plateau formation which extends over
the entire interval $1 \leq \Delta t \leq 9$, for the smallest sea quark mass.
This precocious ground state dominance in the \ep ~channel provides us with
hitherto unobserved accuracy for the \ep -mass estimate, over the entire range
of sea quark masses of the SESAM QCD simulation!

Note from \fig{fig:effmass} that the flavour-nonsinglet mass (marked `$\pi$')
is clearly resolved  from the \ep -mass plateau, with increasing mass gap as
 the quark mass is lowered, at our value of lattice spacing, $a^{-1}(m_{\rho})=
2.30$ GeV.  

Since TEA works better with decreasing sea quark masses and Lanczos methods do
not deteriorate with growing condition numbers, TEA is expected to be
by far superior to SET in the deep chiral regime  of tomorrow's simulations
where linear solvers will tend to become unstable.

\section{Discussion and outlook}

Our  analysis of the \ep~ signal in two-flavour QCD has led us to
the reassuring result that spectral methods in conjunction with ground state
projection on the connected contribution of the \ep -correlator can resolve
the \ep -$\pi$ mass gap before the chiral and continuum extrapolations. In
\fig{fig:extrapol} we show that the lattice data are sufficient to sustain a
chiral extrapolation as well. The figure demonstrates moreover that there is
full consistency of TEA with SET data when being analysed in the same manner.
Before continuum extrapolation, a linear fit of $m_{\eta'}^2(m_{sea}$ works
with reasonable $\chi^2$ and results in $M_{\eta'} = 307(15)$ MeV.  This
value is far away from the actual experimental number, $m_{\eta'} = 957 $
MeV; this does not come as a surprise, however,  since the SESAM QCD vacuum
configurations are missing the strange sea quark contributions.
Therefore our pseudoscalar flavour singlet meson resembles more an (octet) $\eta$ than
 the  \ep ~meson proper. Apart from the continuum extrapolation {\it t.b.d.}, it
is therefore necessary to introduce strange quarks into the calculation, in
order to make contact with real experiment.

Since $N_f = 3$ simulations are not feasible at this time the next step
would be to consider both $\eta $ and \ep ~in a partially quenched
setting with two degenerate light quarks $n$ in the
sea, and a strange valence quarks added. This opens the scenario of a
coupled two channel approach of pseudoscalar singlet mesons
in the $n$ and $s$ sectors, in terms of the 
quark flavour basis~\cite{kroll}
\begin{eqnarray}
|\eta> &=&  \quad \cos \Phi \; |\eta_{\mbox{nn}}>+ \sin \Phi\;  |\eta_{\mbox{ss}}>
\\
|\eta'>& =& -\sin \Phi\; |\eta_{\mbox{nn}}>+ \cos \Phi \;
|\eta_{\mbox{ss}}> \nonumber 
\end{eqnarray}
with \eq{eq:etacorr} being replaced by a $2\times 2$ correlator
\begin{eqnarray}
 C(t) = 
 \left( 
\begin{array}{cc}
 C_{nn}(t) - 2 D_{nn}(t)  & - \sqrt{2}D_{ns}(t) \nonumber\\
 -\sqrt{2}D_{sn}(t)  &  C_{ss}(t)- D_{ss}(t) \nonumber
\end{array}  
\right)\; .
\end{eqnarray}
So far we have dealt only with the left hand upper corner of this matrix. The
evaluation of the OZI-rule violating contributions in its remaining entries
are expected to proceed along the lines  described above, without any
additional complication. The mixing problem then amounts to the solution of
the eigenvalue problem~\cite{luescher}

\beq C(t)C^{-1}(t_0) |i> = \lambda_i(\Delta t) |i> \quad \mbox{with} \; t >
t_0 \eeq with $t-t_0 = \Delta t$.  The two eigenmodes coincide with the
physical \ep ~and $\eta$ states.  Their masses are to be extracted through the
asymptotic relations \beq \lambda_i(\Delta t) \stackrel{\Delta t \to
  \infty}{\longrightarrow} \mbox{exp} (-m_i\Delta t) \; .  \eeq Previously,
the $\eta$-\ep ~ mixing could only be addressed by perturbative
modeling~\cite{ukqcd}.  One would expect that the precocious mass plateau
formation established in the $nn$-sector will persist in the presence of
channel mixing from this eigenvalue problem. Work along this line is in
progress~\cite{dfg}.

{\bf Acknowledgements} K.S.  thanks A.G. Williams, A.C. Kalloniatis, W.
Melnitchouk and their staff for the inspiring atmosphere of the 2001 Cairns
Workshop on Lattice Hadron Physics. A.T. and H.N are  supported  by the
EC Human Potential Programme,  contracts
HPRN-CT-2000-00145 and HPRN-CT-2000-00130.

\end{document}